%% file: main.tex
\documentclass[conference]{IEEEtran}
\usepackage{cite}
\usepackage{amsmath,amssymb,amsfonts}
\usepackage{algorithmic}
\usepackage{graphicx}
\usepackage{textcomp}
\usepackage{xcolor}
\usepackage[hyphens]{url}
\usepackage{subcaption}
\usepackage{pifont}
\usepackage{tcolorbox}
\tcbuselibrary{most}
\usepackage{multirow}
\usepackage[hidelinks]{hyperref}
\usepackage{booktabs}
\usepackage{soul}
\newcommand{\matmul}{\textit{Matmul}}
\newcommand{\prefill}{\textit{prefill}}
\newcommand{\decoding}{\textit{decoding}}

\def\BibTeX{{\rm B\kern-.05em{\sc i\kern-.025em b}\kern-.08em
    T\kern-.1667em\lower.7ex\hbox{E}\kern-.125emX}}

\newcommand{\itemding}[1]{\item[{#1}]}
\newcommand{\allreduce}{\textit{all-reduce}}

\definecolor{nvidiagreen}{HTML}{76B900}

\title{A Hardware Evaluation Framework for Large Language Model Inference}

\author{
Hengrui Zhang, August Ning, Rohan Prabhakar, David Wentzlaff\\
Princeton University, Princeton, New Jersey, USA \\
\{hengrui.zhang, aning, rohanbp, wentzlaf\} @princeton.edu
}

\begin{document}
\maketitle
\thispagestyle{plain}
\pagestyle{plain}


\begin{abstract}

The past year has witnessed the increasing popularity of Large Language Models (LLMs). Their unprecedented scale and associated high hardware cost have impeded their broader adoption, calling for efficient hardware designs. With the large hardware needed to simply run LLM inference, evaluating different hardware designs becomes a new bottleneck.

This work introduces LLMCompass, a hardware evaluation framework for LLM inference workloads. LLMCompass is fast, accurate, versatile, and able to describe and evaluate different hardware designs. LLMCompass includes a mapper to automatically find performance-optimal mapping and scheduling. It also incorporates an area-based cost model to help architects reason about their design choices. 
Compared to real-world hardware, LLMCompass' estimated latency achieves an average 10.4\% error rate across various operators with various input sizes and an average 4.1\% error rate for LLM inference. With LLMCompass, simulating a 4-NVIDIA A100 GPU node running GPT-3 175B inference can be done within 16 minutes on commodity hardware, including 26,400 rounds of the mapper's parameter search.

With the aid of LLMCompass, this work draws architectural implications and explores new cost-effective hardware designs. By reducing the compute capability or replacing High Bandwidth Memory (HBM) with traditional DRAM, these new designs can achieve as much as 3.41x improvement in performance/cost compared to an NVIDIA A100, making them promising choices for democratizing LLMs.

LLMCompass is planned to be fully open-source.

\end{abstract}

\input{text/1-intro}
\input{text/2-preliminary}
\input{text/3-design}
\input{text/6-implication}

\input{text/4-proposal}
\input{text/7-conclusion}


\bibliographystyle{IEEEtranS}
\bibliography{refs}

\end{document}

%% file: text/1-intro.tex
\section{Introduction}

Large Language Models (LLMs), the technology behind OpenAI ChatGPT~\cite{chatgpt}, Github Copilot~\cite{github-copilot}, and Google Bard~\cite{bard}, are gaining widespread attention from the whole society. 
The capability of LLMs is related to their model size~\cite{kaplan2020scaling,hoffmann2022training}, and larger models~\cite{gpt3,chowdhery2022palm} show impressive abilities~\cite{wei2022emergent} 
compared to smaller counterparts~\cite{gpt2,devlin2018bert}, with future models expected to exceed trillions of parameters~\cite{switch-transformer}.

This unprecedented scale of LLMs poses challenges to deployment. 
Serving a GPT-3 (175B parameters) inference requires a minimum of five NVIDIA A100s solely to accommodate the model parameters (in half precision). This substantial hardware cost impedes the broader adoption of LLMs and motivates computer architects to design more cost-effective hardware.
We identify three challenges that exist in designing hardware for LLM inference:

\textbf{Lack of tools to evaluate hardware designs.} Before diving into writing the RTL code, hardware designers may want to first sketch and compare different design choices. There are many properties we want for such a hardware evaluation tool before writing RTL. 
\ding{172}~\textbf{Fast and accurate.} 
Due to the intense compute and memory hardware demand required for LLM inference, 
this tool needs to be as fast as possible without sacrificing accuracy.
\ding{173}~\textbf{Architecturally descriptive.} This tool should be general enough to describe different design choices: If it only applies to a specific architecture, the design space for computer architects will be limited.
\ding{174}~\textbf{Performance-optimal.} The hardware performance is also affected by how the software is programmed (\textit{e.g.}, how to map the workload to the hardware). The evaluation tool should optimize this software domain to fully demonstrate the hardware capability of each design.
\ding{175}~\textbf{Cost-aware.} We also want to know how different hardware design choices affect the hardware cost to reason about cost-performance trade-offs.

Existing tools fail to meet these requirements. Roofline model analysis is fast but not accurate, and cycle-level simulators are accurate but slow. FPGA emulation is accurate and provides area statistics but requires significant engineering effort. To evaluate large-scale hardware designs in the era of LLMs, a new hardware evaluation tool is needed.

\begin{figure}[!t]
    \centering
    \includegraphics[width=0.47\textwidth]{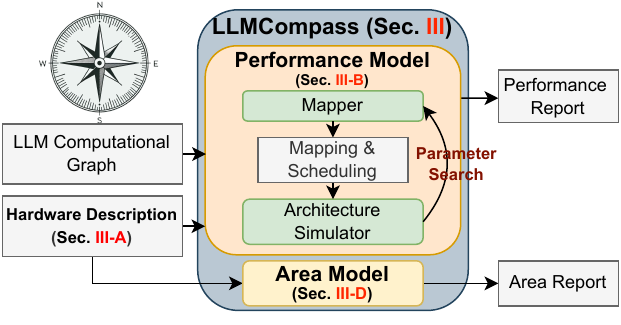}
    \caption{An Overview of LLMCompass. LLMCompass can aid the hardware design process as a versatile evaluation tool.}
    \label{fig:overview}
    \vspace{-5pt}
\end{figure}

\textbf{Lack of knowledge on how different hardware design choices affect LLM inference performance.}
As an emerging application, the hardware characteristics of LLMs remain to be understood. Besides the large volume of compute and memory requirements, LLMs are also unique in their auto-regressive way of generating tokens. We are interested in exploring whether these properties of LLMs will change common architecture wisdom.

\textbf{Lack of cost-effective hardware designs to democratize LLMs.} LLMs are powerful and capable, but are cost-prohibitive to deploy. 
To serve GPT-3, a DGX A100 compute node can cost over \$100,000 USD~\cite{dgx-a100-press-release}, with each NVIDIA A100 featuring 54B transistors and 80 GB of High Bandwidth Memory (HBM).
This high hardware cost hinders democratizing LLMs.

In this paper, we tackle these challenges and make three main contributions.

\textbf{(1) We introduce LLMCompass, a hardware evaluation framework for LLM inference workloads (Sec.~\ref{sec:llmcompass}).} 
LLMCompass leverages the fact that mainstream ML hardware platforms share many architectural commonalities, allowing us to develop a general hardware description template for them.
We also observe LLMs' computational graphs are composed of dense operators: matrix multiplication, softmax, layer normalization, \textit{etc.}, all of which have a structural and hence predictable compute and memory access pattern.
This allows LLMCompass to perform faster, higher-level tile-by-tile (block-by-block) simulations without losing accuracy compared to cycle-accurate simulators.
The framework implements a mapper to manually manage the memory hierarchy and find the performance-optimal mapping and schedule scheme for dense workloads. 
LLMCompass also features a cost and area model based on public parameters to help designers reason about different design choices. 

LLMCompass is validated on three commercial hardware designs: NVIDIA A100~\cite{a100_whitepaper}, AMD MI210~\cite{mi250-whitepaper}, and Google TPUv3~\cite{tpuv2-tpuv3-design-process-ieeemicro2021, tpu-training-cacm2020}. 
Compared to real-world hardware, LLMCompass' estimated latency achieves 10.4\% error rate across various operators with various input sizes and 4.1\% error rate for LLM inference. 
Implemented in Python, LLMCompass is still fast. It takes only 15-16 minutes to simulate a 4-A100 GPU node running GPT-3 175B inference, including 26,400 rounds of the mapper's parameter search (Figure~\ref{fig:perf model validation:gpu prefill}, tested on one core of Intel Xeon Gold 6242R CPU @ 3.10GHz).

\textbf{(2) We leverage LLMCompass to draw architectural implications and explore how hardware design choices affect LLM inference (Sec.~\ref{sec:implication}).} 
We find that \prefill{} and \decoding{} pose different hardware requirements. \textit{Prefill} can significantly benefit from more compute capability and buffers, while \textit{decoding} barely gains from these and is more sensitive to memory bandwidth. These insights inspire us to think about new hardware design paradigms.

\textbf{(3) We propose two cost-effective hardware designs different from conventional wisdom (Sec.~\ref{sec:proposal}).} We find that today's hardware design paradigms tend to fit massive compute capability and SRAMs in a huge die connected to high-end HBMs. We analyze the LLM inference characteristics and show how current hardware designs are inefficient.
\ding{172}~As LLM inference is mostly IO-bound, HBMs can be used to achieve low latency. However, HBM memory capacity limits the batch size, making it hard to fully utilize the massive compute capability. Based on this observation, we find that 95.3\% of the original performance can still be achieved even if we prune the compute capability and buffer size by half.
\ding{173}~Larger batch size can significantly improve throughput as the model parameters are only read once for the whole batch. As memory capacity limits the batch size therefore limiting throughput, we propose to replace HBMs with traditional DRAM. We find that a larger batch size can compensate for the loss in memory bandwidth and can bring a 1.42x improvement in throughput and a 3.41x improvement in performance/cost. 

%% file: text/2-preliminary.tex
\section{Background}

\subsection{Large Language Models and Transformers}

Large Language Models are variations of Transformer models~\cite{vaswani2023attention} with a considerable amount of parameters that have been pre-trained on large corpora of data~\cite{llm_survey}.
Today's LLMs can have as much as one trillion parameters~\cite{switch-transformer}. Compared to smaller models, larger models (\textit{e.g.}~GPT-3 175B~\cite{gpt3}) showcase a remarkable set of capabilities such as emergent abilities~\cite{wei2022emergent} and few-shot learning~\cite{gpt3}.
This increase in model size and the consequent memory and compute requirements have posed unique challenges for hardware.


\begin{figure}[!t]
    \centering
    \includegraphics[width=0.43\textwidth]{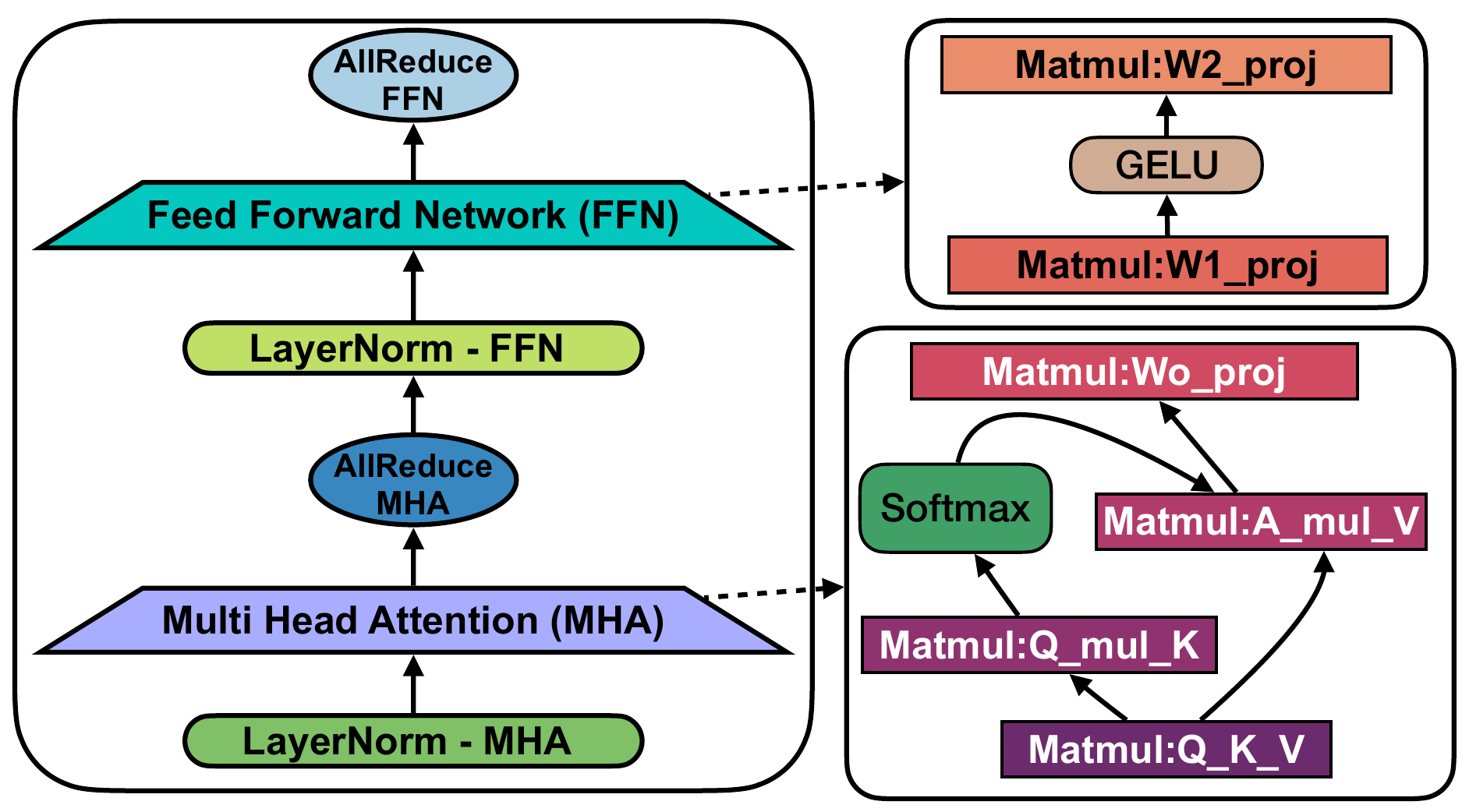}
    \caption{A Decoder-Only Transformer Layer with Tensor Parallelism. GPT-3 175B~\cite{gpt3} consists of a stack of 96 such layers.}
    \vspace{-7pt}
    \label{fig:transformer_layer}
\end{figure}


\begin{figure*}[!t]
    \centering
    \includegraphics[width=0.9\textwidth]{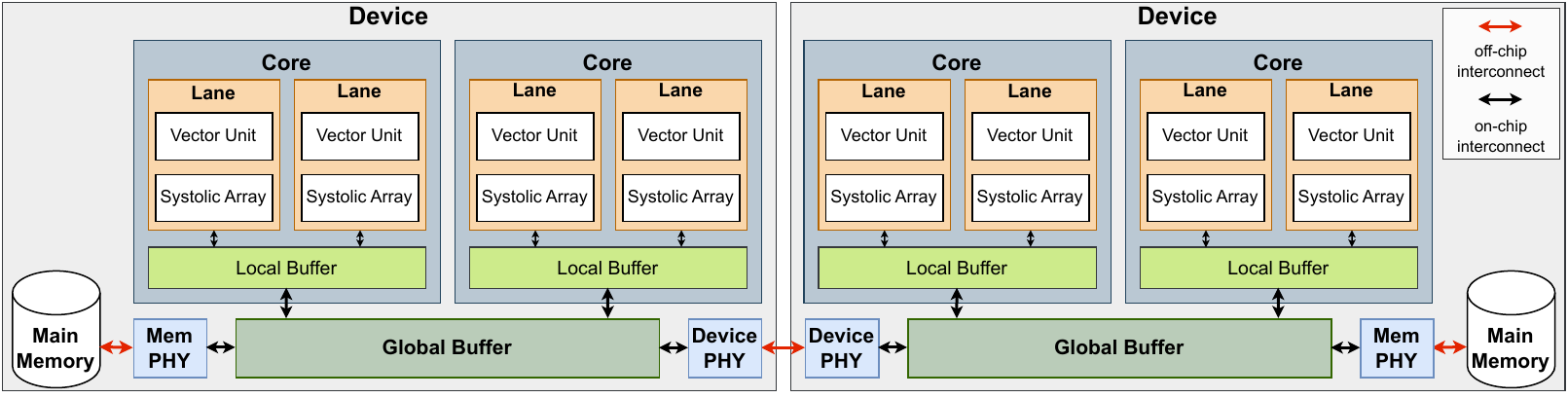}
    \caption{LLMCompass' Hardware Description Template. In this example, each device has 2 cores and each core has 2 lanes.}
    \vspace{-5pt}
    \label{fig:arch abstraction}
\end{figure*}

We focus on Decoder-only Transformer models~\cite{formal_algorithms}, which is the architecture adopted by most of the LLMs today: LLaMA~\cite{touvron2023llama}, GPTs~\cite{gpt3,gpt2}, Bloom~\cite{workshop2023bloom}, PaLM~\cite{chowdhery2022palm},~\textit{etc.} The basic building blocks of these models are Transformer layers. As illustrated in Figure~\ref{fig:transformer_layer}, each layer comprises a Multi-Head Attention block followed by an MLP block.
These layers are then stacked together, forming the bulk of an LLM's memory and compute requirement.
Transformers also use learned Vocabulary and Position embeddings, but for large models like GPT-3, these do not contribute significantly to either the memory or compute requirement ($<2$\%).
Without losing generality, we focus on Multi-Head Attention Transformers (GPT-style). There are other variations such as Multi-Query Attention~\cite{chowdhery2022palm}, Mixture-of-Experts~\cite{switch-transformer}, and parallel Attention and MLP~\cite{chowdhery2022palm}.
LLMCompass seamlessly supports all these possible variations as they share a common set of operators.

\subsection{LLM Inference}

Given an input prompt and the required number of output tokens, LLM inference can be divided into two stages~\cite{pope2023efficiently}.
\ding{172} \textit{Prefill}: Processing the input prompt and computing the KV cache. The Key Value (KV) cache refers to the stored Key and Value tensors of the Attention block in each layer~\cite{pope2023efficiently}. 
\ding{173} \textit{Decoding}: Generating output tokens one after another in an auto-regressive manner: The Key and Value of the newly generated token will be concatenated to the KV cache and used for generating the next token.
The latency of \textit{prefill} and \textit{decoding} is mostly determined by the input and output sequence lengths, respectively.
In \textit{prefill}, as the entire input sequence needs to be multiplied by all the parameters, it is usually bounded by compute.
In \textit{decoding}, each new token needs to be multiplied by all the parameters and concatenated to the KV cache, so \textit{decoding} is usually bounded by reading parameters and KV cache.

Latency and throughput are the key metrics to evaluate LLM inference systems.
For interactive use cases such as chatbots~\cite{chatgpt}, it is imperative to optimize latency.
For background data processing use cases such as data wrangling~\cite{datawrangle} or form processing~\cite{chen2021spreadsheetcoder}, throughput is more important.
The tradeoff between latency and throughput is determined by batch size: larger batch increases throughput at the cost of higher latency.

\subsection{Parallelizing LLM Inference}

Due to the large volume of compute and memory operations, it is beneficial to parallelize LLM inference across multiple devices.
This leads to much better performance and can be necessary if the model's parameters along with the KV cache do not fit in a single device's memory.
For LLM inference, there are two model parallelization schemes: pipeline parallelism and tensor parallelism.
In pipeline parallelism, different layers of the model are grouped into sequential partitions and assigned to different devices like a hardware pipeline.
This scheme has the effect of considerably increasing throughput at the expense of increased latency.
On the other hand, tensor parallelism, as proposed by Megatron-LM~\cite{megatron_lm}, partitions each layer of the model across the available devices, thereby decreasing latency at the cost of frequent device-device communication and synchronization.
As shown in Figure~\ref{fig:transformer_layer}, this scheme requires two \allreduce{} for each Transformer layer, one after the Attention block and another after the MLP block.


%% file: text/3-design.tex
\section{LLMCompass}\label{sec:llmcompass}

An overview of LLMCompass (\textbf{L}arge \textbf{L}anguage \textbf{M}odel \textbf{Com}putation \textbf{P}erformance and \textbf{A}rea \textbf{S}ynthesi\textbf{s}) is shown in Figure~\ref{fig:overview}. To evaluate the performance (\textit{e.g.}, throughput and latency) of running a Transformer-based large language model on a hardware system, two inputs are needed: the computational graph of the LLM and a \textbf{hardware description}~(Section~\ref{sec:arch abstraction}). Given the input, the \textbf{performance model}~(Section~\ref{sec:perf model}) generates a performance report. The \textbf{mapper} conducts a parameter search along with the \textbf{architecture simulator} to find the best mapping and scheduling scheme. At the same time, the \textbf{area model}~(Section~\ref{sec:area model}) generates the area and cost report.

\subsection{Hardware Description Template}\label{sec:arch abstraction}

The hardware description template of LLMCompass is introduced below, as shown in Figure~\ref{fig:arch abstraction}:
\begin{itemize}
\item A \textbf{system} (\textit{e.g.}, a DGX node) is composed of multiple devices connected through a device-device interconnect (\textit{e.g.}, NVLink or Infinity Link). 
\item Each \textbf{device} (\textit{e.g.}, a GPU) is composed of multiple cores, a shared global buffer, and an off-chip main memory. The \textbf{global buffer} (\textit{e.g.}, L2 cache in NVIDIA GPUs) is connected to the main memory, device-device interconnect, and all the cores.
\item Each \textbf{core} (\textit{e.g.}, a Stream Multiprocessor in NVIDIA GPUs) can have multiple lanes sharing a \textbf{local buffer} (\textit{e.g.}, L1 cache in NVIDIA GPUs). The local buffer is connected to the global buffer through the on-chip interconnect.
\item Each \textbf{lane} is independent from each other and has its own \textbf{vector unit}, \textbf{systolic array}, registers and control logic.
\end{itemize}

\begin{table}[!t]
    \centering
    \caption{Examples of LLMCompass's Hardware Description}
    \resizebox{0.48\textwidth}{!}
    {
    \begin{tabular}{c|ccc}
    \toprule
    \multirow{2}{*}{\textbf{Key Specifications}} & \textbf{NVIDIA} & {\textbf{AMD}} & \textbf{Google} \\
    & \textbf{A100~\cite{a100_whitepaper}} & \textbf{MI210~\cite{mi250-whitepaper}} & \textbf{TPUv3\protect\footnotemark~\cite{tpuv2-tpuv3-design-process-ieeemicro2021}} \\
    \midrule
    Frequency (MHz) & 1410 & 1700 & 940\\
    Core count&{108}& 104&2\\
    Lane count& 4 & 4 &1\\
    Vector width & 32 & 16 & $4\times128$\\
    Systolic array & $16\times16$ & $16\times16$ & $128\times128$\\
    Local buffer (KB)& 192 & 80 & 8192\\
    Global buffer (MB)& {40} & 8 & 16384\\
    {Global buffer (bytes/clk)} & {5120} & {{4096}} & 490\\
    Memory bandwidth (TB/s) & 2 & 1.6 & - \\
    Memory capacity (GB) & 80 & 64 & - \\
    Device-device bandwidth (GB/s) & 600 & 300 & 162.5\\
    \bottomrule 
    \end{tabular}
    }
    \label{tab:example}
\end{table}
\footnotetext{One TPUv3 core. Each TPUv3 chip has two TPUv3 cores. TPUv3 cores within the same chip are connected by internal links.}

\begin{figure*}[!t]
    \centering
    \includegraphics[width=0.93\textwidth]{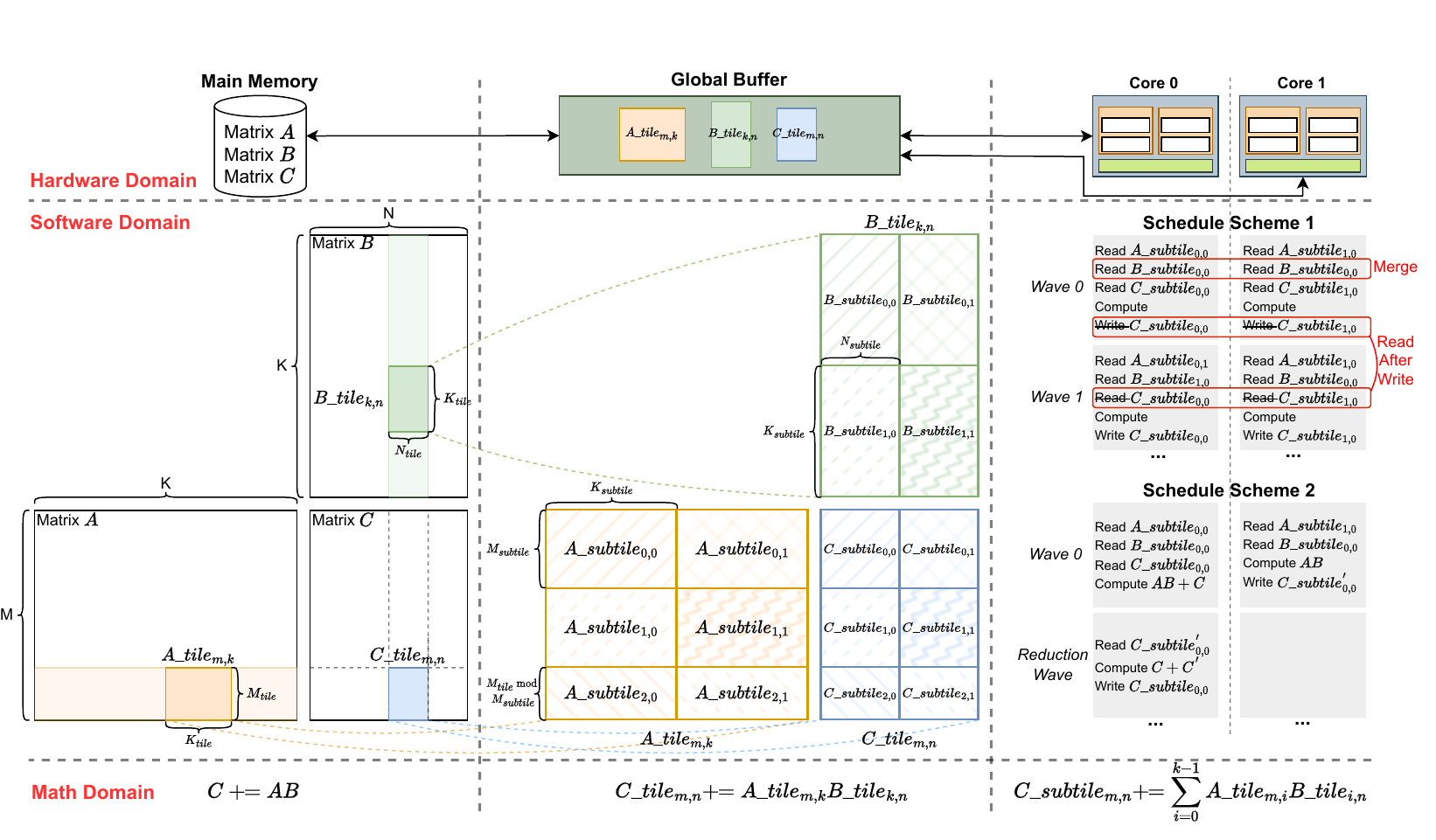}
    \vspace{-5pt}
    \caption{Visualization of a Matrix Multiplication in LLMCompass as in Section~\ref{sec:matmul}.}
    \vspace{-5pt}
    \label{fig:matmul}
\end{figure*}

In existing devices, the local and global buffers are usually on-chip SRAM: cache, scratchpad, or a combination of both. LLMCompass doesn't distinguish between cache and scratchpad because the memory is explicitly managed by the mapper. We believe this assumption does not lose generality as a highly optimized library will also carefully manage the memory. The main memory is usually off-chip DRAM: HBM, DDR memory, CXL memory,~\textit{etc}, all of which can be described by our parameterized hardware description template.

We find this hardware description is general enough to describe the mainstream machine learning platforms of today: NVIDIA GPUs, AMD GPUs, and Google TPUs, as shown in Table~\ref{tab:example} with a sample of key specifications listed. It is also flexible enough to explore future architectures.



\subsection{Performance Model}\label{sec:perf model}

The computational graph of a Transformer is composed of a stack of Transformer layers.
Each layer is composed of a series of operators, including matrix multiplication (\textit{Matmul}), \textit{Softmax}, layer normalization (\textit{LayerNorm}), and activation functions (\textit{e.g.}, GELU~\cite{gelu} as in GPTs~\cite{gpt3,gpt2}). In a multi-device setup, communication primitives such as \textit{all-reduce} operators are also needed to perform tensor parallelism. The key challenge is how to simulate the performance of different operators and communication primitives on a given hardware system - this requires knowledge about the hardware and how to map and schedule operators on a multi-level compute system with a multi-level memory hierarchy.

To solve this, LLMCompass introduces a mapper and an architecture simulator to build a performance model. Conceptually, we simulate running an operator on the chosen hardware in a recursive manner: we first partition the problem into smaller sub-problems that can fit in the global buffer. The sub-problem is then divided into smaller sub-sub-problems that can fit in each core's local buffer. The partitioning, mapping, and scheduling are generated by the mapper and a parameter search is conducted to find the optimal mapping and scheduling. LLMCompass always tries to find the performance-optimal mapping to fully demonstrate the hardware capability.

\subsubsection{\textbf{Matrix Multiplication}}\label{sec:matmul}

The process of simulating a matrix multiplication is visualized in Figure~\ref{fig:matmul}. $\mathbf{A}$ is a $M\times K$ matrix with $M$ rows and $K$ columns. Similarly, $\mathbf{B}$ and $\mathbf{C}$ are $K\times N$ and $M\times N$ matrices respectively. A generalized matrix multiplication is defined as $\mathbf{C}= \mathbf{AB}+\mathbf{C}$.

\textbf{From main memory to global buffer:} To maximize data reuse, matrix multiplication is usually 
calculated in a tile-by-tile manner~\cite{tiling}. As shown on the left of Figure~\ref{fig:matmul}, matrix $\mathbf{A}$, $\mathbf{B}$, and $\mathbf{C}$ are divided into tiles small enough to fit into the global buffer. In each step, one $A\_{tile}_{m,k}$, $B\_{tile}_{k,n}$, and $C\_tile_{m,n}$ are read into the global buffer, the cores then perform the computation, and the results are written back.

\textbf{From global buffer to local buffer:} With tiles inside the global buffer, we now need to parallelize the computation of $C\_tile_{m,n}=A\_{tile}_{m,k}B\_{tile}_{k,n}+C\_tile_{m,n}$ on multiple cores. As shown in the middle of Figure~\ref{fig:matmul}, these tiles are further divided into smaller sub-tiles to fit in each core's local buffer. It then becomes a scheduling problem to map sub-tiles onto cores.

The right of Figure~\ref{fig:matmul} shows two possible schedule schemes: 
\begin{itemize}
\item \textbf{Schedule Scheme 1:} Different cores working on different $C\_subtile$s in the same column. At \textit{wave~0}, as \textit{core~0} and \textit{core~1} both need to read the same $B\_subtile$, their memory access to the global buffer should be merged. In our simulator, this memory access merging is automatically identified and taken care of. As the same core keeps updating the same $C\_subtile$, there is no need to first write the partial result and then read it from the global buffer. This \textit{Read-After-Write} dependency is also automatically taken care of by the simulator.
\item \textbf{Schedule Scheme 2:} Different cores working on the same $C\_subtile$. \textit{Core~0} and \textit{core~1} first read the data and calculate the partial results, then perform a reduction and write back the final results. 
\end{itemize}

In reality, with more cores and more tiles, the schedule space can be more complicated than the example shown in Figure~\ref{fig:matmul}. 

\textbf{From local buffer to lanes:} 
Similarly, within each core, the sub-tiles are further partitioned into sub-sub-tiles to be mapped to lanes sharing a local buffer. After that, the sub-sub-tiles are finally passed to the systolic arrays. LLMCompass leverages SCALE-Sim~\cite{scale-sim-1,scale-sim-2}, a cycle-level systolic array simulator, to mimic the behavior of a systolic array and get the cycle count. LLCompass caches the results of SCALE-Sim into a look-up table to avoid duplicated simulation. A reduction will be performed by the vector unit if needed.

\textbf{Mapper:} A parameter search is performed by the mapper to determine the best tiling scheme and schedule scheme. To overlap computation with memory accesses, we also add software pipelines (double buffering) at each level of the memory hierarchy as scheduling options. The downside of enabling software pipeline is that it requires extra buffer space so the maximal tile size will be reduced, causing potentially lower utilization of systolic arrays. However, we find software pipeline to be beneficial in most cases.

\subsubsection{\textbf{Communication Primitives}}

We use the link model as in AHEAD~\cite{ahead} and LogGP~\cite{loggp}. Suppose $L$ is the link latency, $O$ is the additional overhead associated with the data transfer, and $B$ is the link bandwidth. The latency $T$ to transfer $n$ bytes of data through a link is expressed in Equation~\ref{eq:link latency} and~\ref{eq:packet overhead}:
\vspace{-5pt}
\begin{gather}
    T=L+O+\frac{\hat{n}}{B}\label{eq:link latency}\\
    \hat{n} =\left\lceil \frac{n}{{MaxPayload}} \right\rceil * {Flit\_size} + n \label{eq:packet overhead}
\end{gather}
\vspace{-5pt}

On top of this, we implement ring all-reduce~\cite{ring-all-reduce}, which is a bandwidth-optimal all-reduce algorithm. We use a 16-byte $Flit\_size$ and a 256-byte $MaxPayload$  based on NVLinks~\cite{nvlink}. We don't model more communication primitives as LLM inference only requires \textit{all-reduce} for tensor parallelism and \textit{peer-to-peer} for pipeline parallelism.

\subsubsection{\textbf{Other Operators}}
We also model \textit{Softmax}, \textit{LayerNorm}, and \textit{GELU} following a similar methodology as in Section\ref{sec:matmul}. The only differences are: \ding{172} They have fewer dimensions and are therefore simpler: \textit{Softmax} and \textit{LayerNorm} operate on two-dimensional data, and \textit{GELU} operates on one-dimensional data, while \matmul{} operates on three-dimensional data. As each dimension requires tiling and scheduling, the mapper search space is much smaller. \ding{173} They do not use systolic arrays. \textit{Softmax} is implemented with the online algorithm~\cite{onlinesoftmax}. \textit{GELU} is approximated with $tanh$~\cite{gelu}.

\begin{figure*}[!t]
    \centering
    \begin{subfigure}[b]{0.27\textwidth}
        \centering
        \includegraphics[width=\textwidth]{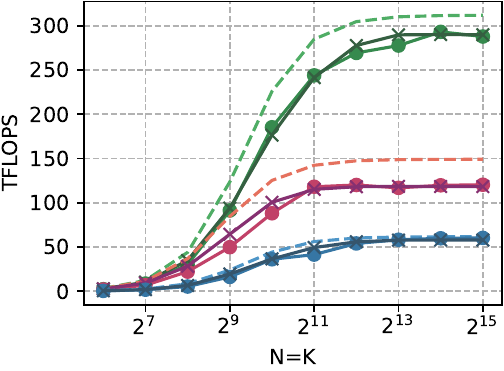}
        \vspace{-15pt}
        \caption{\textit{Matmul} with $M=8192$.}
    \end{subfigure}
    \hfill
    \hspace{-5pt}
    \begin{subfigure}[b]{0.27\textwidth} 
        \centering
        \includegraphics[width=\textwidth]{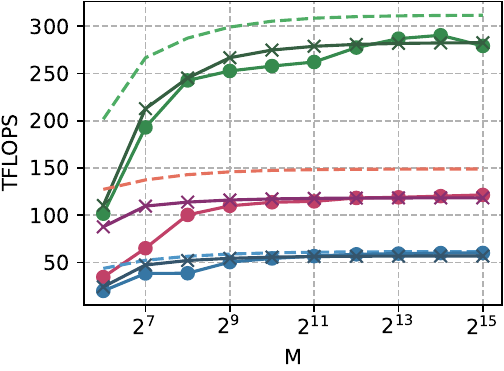}
        \vspace{-15pt}
        \caption{\textit{Matmul} with $N=K=12288$.}
    \end{subfigure}
    \hfill
    \hspace{-5pt}
    \begin{subfigure}[b]{0.445\textwidth}
        \centering
        \includegraphics[width=\textwidth]{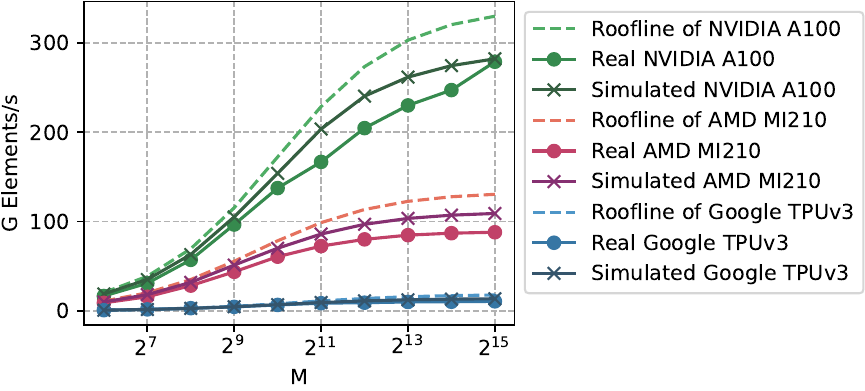}
        \vspace{-15pt}
        \caption{\textit{Softmax} with $N=4096$.\ \ \ \ \ \ \ \ \ \ \ \ \ \ \ \ \ \ \ \ }
    \end{subfigure}

    \begin{subfigure}[b]{0.27\textwidth} 
        \centering
        \includegraphics[width=\textwidth]{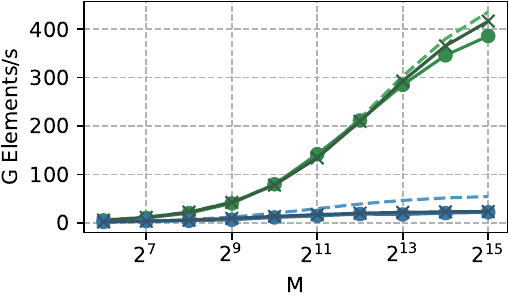}
        \vspace{-15pt}
        \caption{\textit{LayerNorm} with $N=4096$.}
        \label{fig:validatation:layernorm n}
    \end{subfigure}
    \hfill
    \hspace{-10pt}
    \begin{subfigure}[b]{0.27\textwidth}
        \centering
        \includegraphics[width=\textwidth]{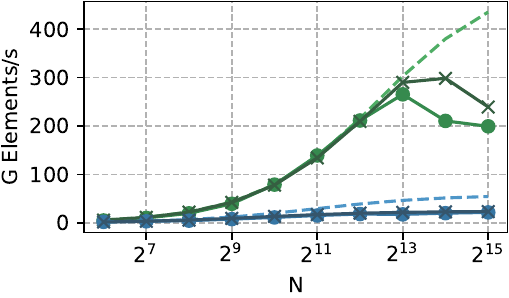}
        \vspace{-15pt}
        \caption{\textit{LayerNorm} with $M=4096$.}
    \end{subfigure}
    \hfill
    \hspace{-10pt}
    \begin{subfigure}[b]{0.45\textwidth}
        \centering
        \includegraphics[width=\textwidth]{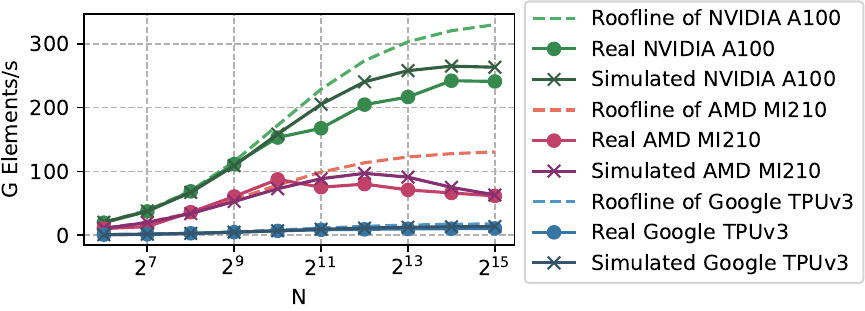}
        \vspace{-15pt}
        \caption{\textit{Softmax} with $M=4096$.\ \ \ \ \ \ \ \ \ \ \ \ \ \ \ \ \ \ \ \ }
    \end{subfigure}

    \begin{subfigure}[b]{0.57\textwidth}
        \centering
        \vspace{5pt}
        \includegraphics[width=\textwidth]{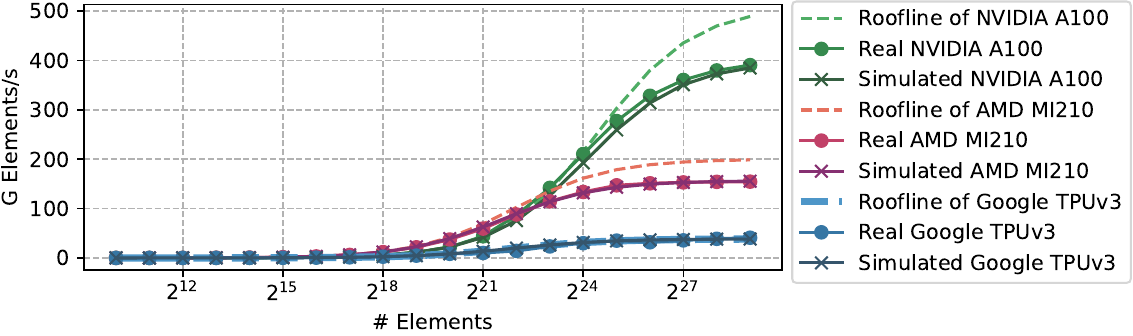}
        \vspace{-15pt}
        \caption{\textit{GELU}.}
    \end{subfigure}
    \hfill
    \begin{subfigure}[b]{0.4\textwidth}
        \centering
        \vspace{5pt}
        \includegraphics[width=\textwidth]{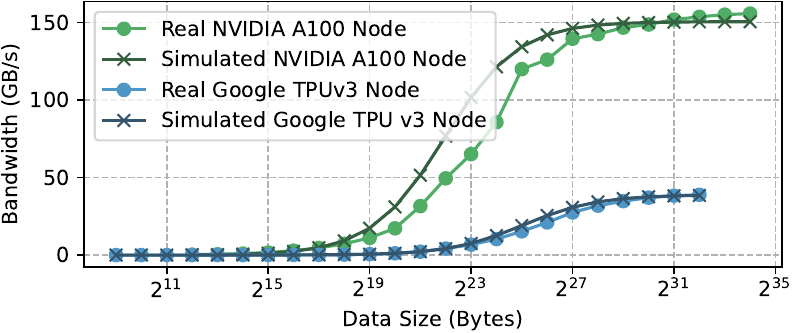}
        \vspace{-15pt}
        \caption{\textit{All-reduce}.}
    \end{subfigure}
    \hspace{30pt}

    \begin{subfigure}[b]{0.2\textwidth}
        \centering
        \vspace{-10pt}
        \includegraphics[width=\textwidth]{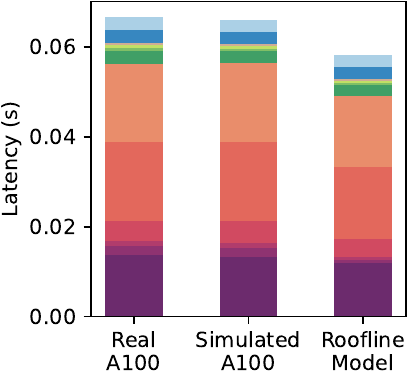}
        \caption{GPU \textit{Prefill} Latency.}\label{fig:perf model validation:gpu prefill}
    \end{subfigure}
    \hfill
    \hspace{-10pt}
    \begin{subfigure}[b]{0.2\textwidth}
        \centering
        \vspace{-10pt}
        \includegraphics[width=\textwidth]{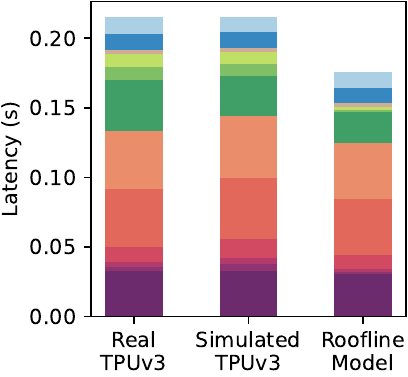}
        \caption{TPU \textit{Prefill} Latency.}
    \end{subfigure}
    \hfill
    \hspace{-10pt}
    \begin{subfigure}[b]{0.2\textwidth}
        \centering
        \vspace{-10pt}
        \includegraphics[width=\textwidth]{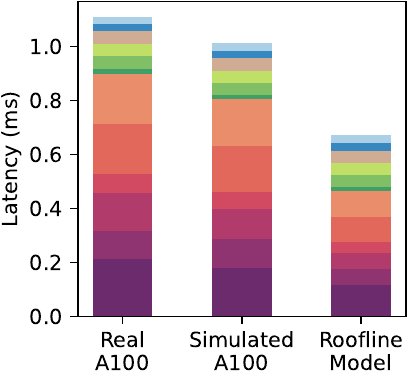}
        \caption{GPU \textit{Decoding} Latency.}
    \end{subfigure}
    \hfill
    \hspace{-10pt}
    \begin{subfigure}[b]{0.31\textwidth}
        \centering
        \vspace{-10pt}
        \includegraphics[width=\textwidth]{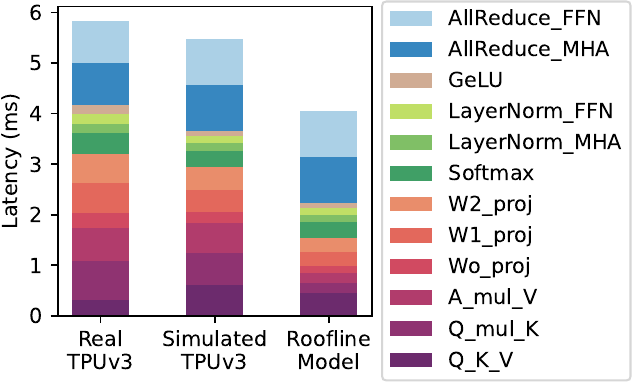}
        \caption{TPU \textit{Decoding} Latency.\ \ \ \ \ \ \ \ \ \ \ \ \ \ \ }
    \end{subfigure}
    \hfill
        
    \caption{Performance Model Validations. \matmul{} takes a $M\times K$ ($M$ rows and $K$ columns) and a $K\times N$ matrix as input. 12288 is the model dimension of GPT-3~\cite{gpt3}. \textit{Softmax} and \textit{LayerNorm} take a $M\times N$ matrix and perform normalization on the $N$ dimension. \textit{Prefill} latency is measured by running one layer of GPT-3 with batch size 8 and sequence length 2048. \textit{Decoding} latency is per GPT-3 layer per output token and is measured by the latency of generating the 1024th output token with batch size 8 and input sequence length 2048. For (a)-(g), a single GPU/TPU device is used. For (h)-(l), the 4-A100 GPU node and 8-TPUv3-Core TPU node are used with tensor parallelism.}
    \label{fig:perf model validation}
    \vspace{-7pt}
\end{figure*}

\subsection{{Performance Model Validation}}

In this section, we validate our framework against three real hardware platforms: (1) a datacenter GPU node with 4 NVIDIA A100 SXM4 GPUs (80 GB) fully connected by NVLinks; (2) a Google Cloud TPU node with 8 TPUv3 cores connected in a 2D torus topology; (3) an AMD MI210 GPU\footnote{We set the frequency to 1400 Mhz to avoid frequency fluctuation}. The results are shown in Figure~\ref{fig:perf model validation}. For NVIDIA GPUs, CUDA 11.7 and PyTorch 2.0 are used to benchmark operators in half precision (FP16) with \texttt{torch.compile} enabled for \textit{LayerNorm} and \textit{GELU} to maximize performance. Communication primitive \textit{all-reduce} is benchmarked with {nccl-tests}~\cite{nccl}, a communication primitive performance benchmark for NVIDIA GPUs. For Google TPUs, JAX 0.4.18 is used to benchmark operators and communication primitives. Due to the hardware feature of TPUs, \textit{Matmul} is benchmarked in bfloat16 (BF16) and all the other operators are in FP32. For AMD GPU, ROCm 5.4.2 and PyTorch 2.0\footnote{We encountered ``load binary'' error so we didn't benchmark \textit{LayerNorm} on AMD MI210.} are used along with FP16 for \textit{Matmul} and FP32 for other operators. The kernel launch overhead including the framework overhead is measured by running the operator with an input of size 1.

As shown in Figure~\ref{fig:perf model validation}, for \textit{Matmul}, \textit{Softmax}, \textit{LayerNorm}, \textit{GELU}, and \textit{all-reduce}, LLMCompass achieves an average error rate of 9.0\%, 12.0\%, 11.3\%, 5.0\%, and 14.9\% respectively.
For LLM inference, LLMCompass achieves an average error rate of 0.69\% and 7.5\% for \textit{prefill} and \textit{decoding} respectively.
\textbf{On average, LLMCompass achieves a 10.4\% error rate for different operators at various input sizes and a 4.1\% error rate across the \prefill{} and \decoding{} stages.
}

Although not a perfect match to real-world hardware, LLMCompass is able to show a similar trend that a naive roofline model fails to show. For example, in Figure~\ref{fig:validatation:layernorm n}, as the reduction dimension of \textit{LayerNorm} increases to an extreme, the throughput should drop due to the increasing reduction cost. LLMCompass is able to catch this trend.

LLMCompass' results are totally interpretable without incorporating any fudge factor and we believe this interpretability is more important than perfectly matched results. Here are some possible causes of the mismatch between LLMCompass and real hardware:
\begin{itemize}
    \item Lack of hardware knowledge. We have little knowledge about the micro-architecture of GPUs and TPUs (\textit{e.g.}, hardware pipeline design or scheduler design). With a large input size, the hardware is well utilized and some overhead can be hidden. However, with a small input size, it's hard to hide the overhead and micro-architecture details affect performance significantly. Also, the Tensor Cores in NVIDIA GPUs and Matrix Cores in AMD GPUs are simulated as systolic arrays in LLMCompass, which may not be true in reality.
    \item Lack of software knowledge. We don't know how operators and communication primitives are implemented on these platforms as they are closed-source libraries. We conduct a thorough parameter search for each input size to maximize performance, but in reality those libraries probably use heuristics to determine mapping and scheduling, which may not be optimal at all input sizes (\textit{e.g.}, we find that for a \textit{Matmul} with $M=64$ and $N=K=12288$, AMD MI210 is less than 25\% of its roofline performance while a NVIDIA A100 can achieve 50\% of its roofline performance.). Also, some key information is not available. For example, we cannot find the packet format for TPU-TPU communication and have to use the NVLink packet format instead.
    \item Non-ideal hardware. LLMCompass assumes a fixed frequency, but when testing real-world hardware, we have no control over the frequency of the datacenter GPU or TPU nodes. LLMCompass also assumes bandwidth can be utilized at full rate, but in reality there may be some other overhead (\textit{e.g.}, error correction code).
\end{itemize}

\input{text/3.1-area}

%% file: text/3.1-area.tex
\subsection{Area and Cost Model}\label{sec:area model}

As chip designers increase die area to improve single chip performance, fewer chips fit per wafer and may also risk decreased yield, leading to increased costs.
LLMCompass incorporates area and cost models to allow designers to reason about these performance-area trade-offs.
These models use the provided hardware description with estimated transistor counts and/or die areas from known components to find the total device die area - our methodology is explained as follows.

Within each core's lanes, we estimate the vector units' and systolic arrays' transistor counts from open-source designs, tape-outs, and generators~\cite{zaruba_ariane_tvlsi2019, mckeown_piton_characterization_hpca2018, genc_gemmini_dac2021}. 
We estimate each lane's register file's area overhead using an empirical area model~\cite{raghavan_empire_2009}.
For the local buffer shared amongst lanes in each core as well as the global buffer shared amongst cores, we model them as SRAM caches and derive their areas using CACTI~\cite{cacti6} and scale results down to a 7nm process. 
For memory and device-device interconnect, we estimate PHY and controller area based on annotated A100 and MI210 die photos~\cite{patel-semianalysis-lovelace-dies-2022, smith_MI200_hotchips2022}. 
In our calculations, the controller area scales based on the process node, but the PHY area remains fixed as they do not scale well due to internal analog devices.

\input{Figures/model/cost_model_params}

\begin{figure}[!t]
    \centering
    \begin{subfigure}[]{0.47\textwidth} 
        \includegraphics[width=0.99\textwidth]{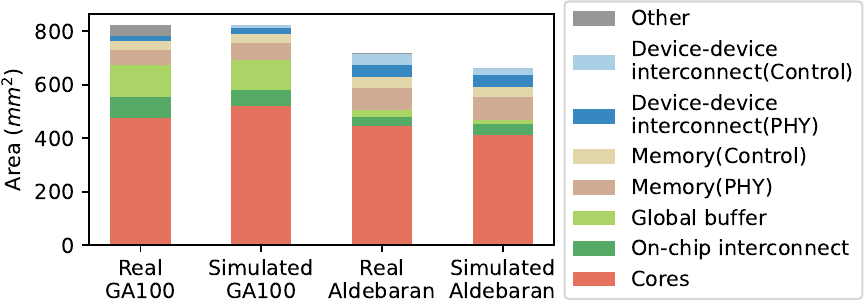}
        \caption{Die Area Breakdown of NVIDIA GA100 and AMD Aldebaran.}
        \vspace{5pt}
        \label{fig:die-area-comparison}
    \end{subfigure}
    \begin{subfigure}[]{0.47\textwidth} 
        \includegraphics[width=0.99\textwidth]{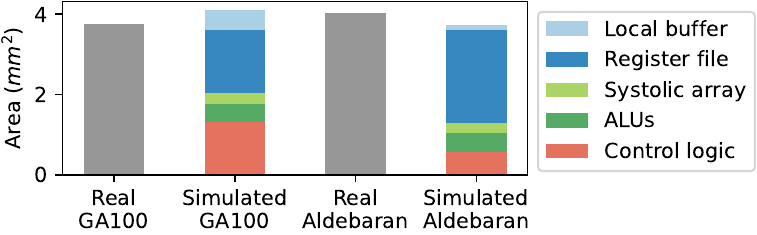}
        \caption{Core Area Breakdown (Stream Multiprocessor for NVIDIA GPUs and Compute Unit for AMD GPUs).}
        \label{fig:core-area-breakdown}
    \end{subfigure}
    
    \caption{Area Model Validations.}
    \vspace{-5pt}
\end{figure}

We account for extra per lane overheads ({\it e.g.}, control signals) by calculating the core area using our model and taking the difference from the expected die areas taken from annotated photos.
We then divide the overhead per lane, per scheduler width (32 in A100s, 16 in MI210).
Similarly, we account for extra per core overheads ({\it e.g.}, core-to-core crossbars) by calculating the expected die area with our model and splitting the area between the cores.
These per-lane and per-core overhead estimates are averaged between AMD and NVIDIA chips.

To estimate cost, LLMCompass uses supply chain modeling~\cite{ning-supply-chain-isca2023} for wafer costs to calculate per-die costs. 
These per-die costs do not incorporate any IP, masks, or packaging costs. 
For memory costs, we use average DRAM spot prices for DDR~\cite{dram-spot-price} and consumer estimates for HBM2e~\cite{hbm-pricing-semieng2019}.

Table~\ref{tab:cost_model_params} shows a sample of the transistor counts and corresponding 7nm die areas of the parameters used in the area model. 
Using their respective architecture white papers, we model GA100~\cite{a100_whitepaper} (the die used in NVIDIA A100) and Aldebaran~\cite{mi250-whitepaper} (the die used in AMD MI210) dies to estimate their total die areas, shown in Fig.~\ref{fig:die-area-comparison}.
For the accounted-for components, LLMCompass' area model estimates for GA100 and Aldebaran dies achieve a 5.1\% and 8.1\% error respectively.
We attribute these differences to the core's microarchitecture and core-to-core communication overheads which are proprietary and difficult to estimate.
Our model also allows users to break down a single core's area into its individual components, shown in Fig.~\ref{fig:core-area-breakdown}.

%% file: Figures/model/cost_model_params.tex
\begin{table}[!t]
    \caption{A Sample of Area Model Parameters (7nm)}
    \centering
    \resizebox{0.47\textwidth}{!}
    {
    \begin{tabular}{c|c|c}
    \toprule
    \textbf{Parameter}  & \textbf{Transistor Count} & \textbf{7nm Area ($\mu{}m^2$)}\\
    \midrule
    64 Bit Floating Point Unit & 685300 & 7116 \\
    32 Bit Int ALU & 177000 & 1838 \\
    Per Lane Overhead & 996200 & 10344 \\
    Per Core Overhead & 44300000 & 460000 \\
    1024 Bit HBM2e Control & 552743000 & 5740000 \\
    1024 Bit HBM2e PHY & - & 10450000 \\
    \bottomrule 
    \end{tabular}
    }
    \label{tab:cost_model_params}
    \vspace{-8pt}
\end{table}

%% file: text/6-implication.tex
\section{Architectural Implications}\label{sec:implication}


With LLMCompass, we are able to conduct a design space exploration and shed light on how to design efficient hardware systems for LLM inference. 
In this section, we use LLMCompass to study how different compute system configurations, memory bandwidth, and buffer sizes affect LLM inference performance and draw architectural implications. These insights inspire us to propose new designs as in Section~\ref{sec:proposal}.


\subsection{Experimental Setup}
For all the unmentioned specifications, we use the specifications of an NVIDIA \textcolor{nvidiagreen}{A100} (as in Table~\ref{tab:example}) and 4-way tensor parallelism. \textit{Prefill} latency is measured by running one GPT-3 layer with batch size 8 (a balancing point between latency and throughput) and input sequence length 2048 (a medium-long sequence for GPT-3). \textit{Decoding} latency is measured as the latency of generating the 1024th output token when running one GPT-3 layer with batch size 8 and input sequence length 2048. We use FP16 for all the operators.

\subsection{Compute System}

We test five different compute system designs as shown in Table~\ref{tab:core size}. From A to E, we increase each core's systolic array, vector unit, and local buffer capacities. B represents a full \textcolor{nvidiagreen}{GA100}. We keep B, C, D, and E to have the same total compute capability and total buffer size to compare the design choice of fewer big cores or more tiny cores. Configuration A only has a quarter of the compute capability compared to others. All the designs have the same amount of total buffer size and register file size scales with vector width.

\begin{figure}[!t]
    \centering
    \begin{subfigure}[]{0.46\textwidth} 
        \includegraphics[width=0.95\textwidth]{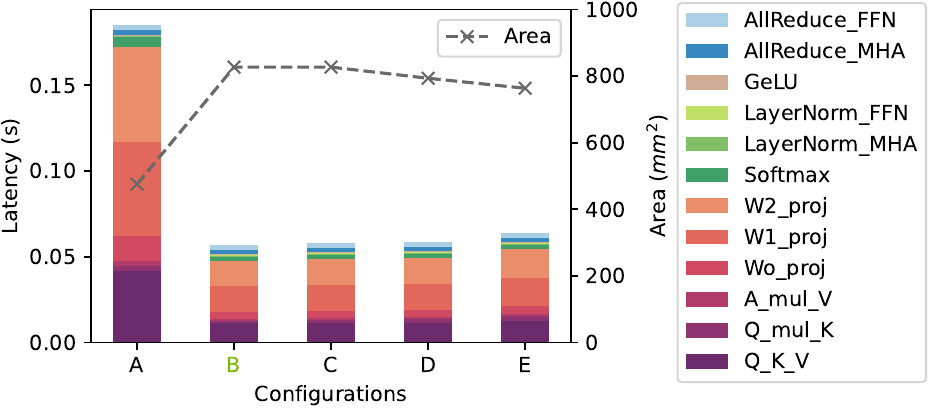}
        \caption{\textit{Prefill} Latency per GPT-3 Layer.}
        \vspace{5pt}
        \label{fig:compute prefilling}
    \end{subfigure}
    \begin{subfigure}[]{0.46\textwidth} 
        \includegraphics[width=0.95\textwidth]{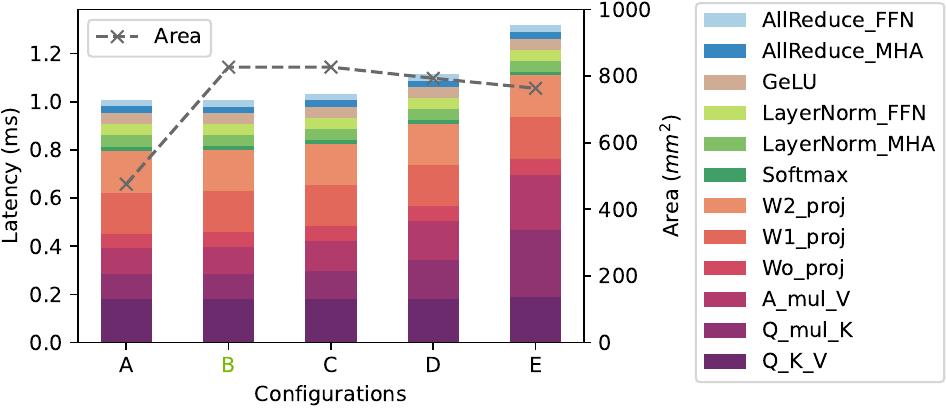}
        \caption{\textit{Decoding} Latency per GPT-3 Layer per Output Token.}
        \label{fig:compute generation}
    \end{subfigure}
    \caption{Impact of Compute System Design on Performance.}
    \vspace{-8pt}
    \label{fig:compute}
\end{figure}

Figure~\ref{fig:compute} shows \prefill{} and \decoding{} latencies for these designs. 
Compared to the \textcolor{nvidiagreen}{GA100}, design A has 3.25x higher \prefill{} latency but is only 0.1\% slower at \decoding{} and uses only 57.8\% of the area. 
Design E with the largest cores see \prefill{} and \decoding{} latency increase by 12.4\% and 30.8\% respectively, but can reduce die area up to 7.7\%.


\textbf{Analysis:} For the \prefill{} stage, B is much faster than A because \prefill{} is compute-bound. As per core systolic arrays and vector units scale, the tile size needs to increase to fully utilize larger computing units. Bigger tiles can cause more padding as the problem size needs to be quantized to the tile size and hardware size. Although large systolic arrays and vector units can be more area-efficient, they are harder to schedule and fully utilize.

Since \decoding{} is IO-bound, increasing compute capability barely helps, which explains why A and B have similar performance. As the matrix multiplications during \decoding{} are narrow (\textit{e.g.} $16 \times{} 12288$), it is even harder to fully utilize larger systolic arrays and performance degrades. 

\begin{tcolorbox}[rounded corners, boxrule=1pt]
    \textit{\textbf{Implications:} }
    \begin{itemize}
    \itemding{\ding{172}} \textit{Increasing compute capability significantly helps \prefill{} but barely helps \decoding{}.}
    \itemding{\ding{173}} \textit{Large systolic arrays are more efficient for \prefill{} compared to \decoding{}.}
\end{itemize}
\end{tcolorbox}

\begin{table}[!t]
    \caption{Five Compute System Designs.}
    \centering
    \resizebox{0.48\textwidth}{!}
    {
    \begin{tabular}{c|ccccc}
    \toprule
    \textbf{Specifications}  & \textbf{A} & {\textbf{B}} & \textbf{C} & \textbf{D} & \textbf{E}\\
    \midrule
    Core count & 128 & 128 & 128& 32&8\\
    Lane count & 4 & 4 & 1 & 1 &1 \\
    Vector width & 8&32&128& 512&2048\\
    Systolic array & $8\times8$ & $16\times16$ & $ 32\times 32$ & $64\times64$& $128\times128$\\
    Local buffer (KB)& 192 &192 &192 &768&3072\\
    \bottomrule 
    \end{tabular}
    }
    \label{tab:core size}
    \vspace{-7pt}
\end{table}

\subsection{Main Memory}

\begin{figure}[!t]
    \centering
    \begin{subfigure}[]{0.45\textwidth} 
        \includegraphics[width=0.99\textwidth]{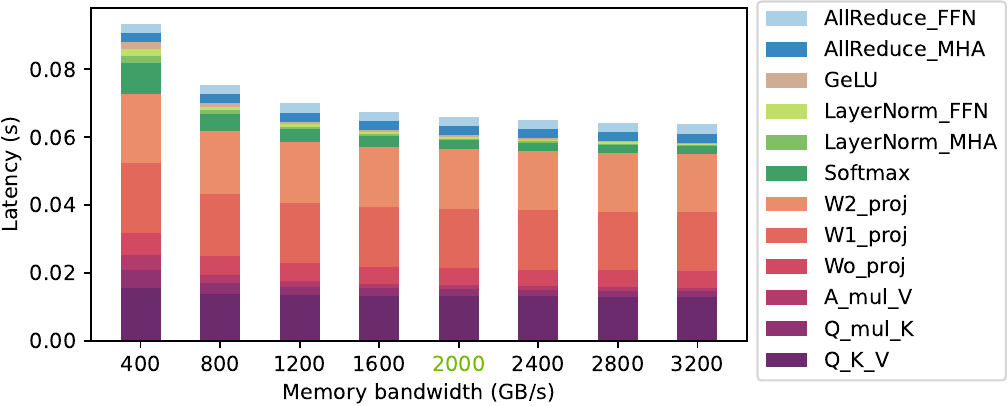}
        \vspace{-2pt}
        \caption{\textit{Prefill} Latency per GPT-3 Layer.}
        \vspace{3pt}
        \label{fig: main memory bw prefilling}
    \end{subfigure}
    \begin{subfigure}[]{0.45\textwidth} 
        \includegraphics[width=0.99\textwidth]{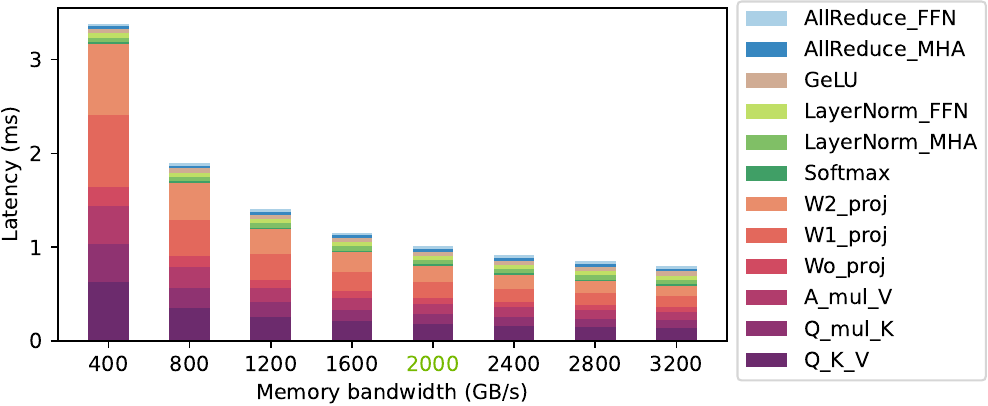}
        \vspace{-2pt}
        \caption{\textit{Decoding} Latency per GPT-3 Layer per Output Token.}
        \label{fig: main memory bw generation}
        \vspace{-3pt}
    \end{subfigure}
    
    \caption{Impact of Memory Bandwidth on Performance.}
    \label{fig:main memory bw}
    \vspace{-7pt}
\end{figure}

As main memory capacity is considered more of a constraint (enough capacity is required to hold the parameters and KV cache), we will focus on the impact of main memory bandwidth. 
Figure~\ref{fig:main memory bw} details the performance results for sweeping memory bandwidth from 400 to 3200 GB/s. 
For \prefill{}, increasing memory bandwidth from 800GB/s to 2000GB/s reduces latency by 14.3\%, and further increasing to 3200GB/s has a marginal performance gain of 3.5\%. For \decoding{}, increasing from 800GB/s to 2000GB/s has a speedup of 1.88x, and further increasing to 3200GB/s brings another 26\% gain.

\textbf{Analysis:} In the \prefill{} stage, \textit{Matmul}s are significantly faster when increasing memory bandwidth from 400GB/s to 800GB/s. Further increasing bandwidth does not significantly affect \textit{Matmul} performance as it becomes compute-bound. For IO-bound \textit{GELU}, \textit{LayerNorm}, and \textit{Softmax}, larger memory bandwidth realizes significant speedup.

In the \decoding{} stage, \textit{Matmul}s are significantly faster with increased memory bandwidth, mainly because they are narrow (turn into a vector-matrix multiplication at batch size 1) and IO-bound. In this stage, \textit{GELU}, \textit{LayerNorm}, and \textit{Softmax} have a small input size. They are dominated by kernel launch overhead and barely affected by memory bandwidth.

\begin{tcolorbox}[rounded corners, boxrule=1pt]
    \begin{itemize}
    \itemding{\ding{174}} \textit{\textit{Decoding} is much more sensitive to memory bandwidth than \prefill{}.}
\end{itemize}
\end{tcolorbox}

\subsection{Local and Global Buffer}

\begin{figure}[t]
    \centering
    \begin{subfigure}[]{0.46\textwidth} 
        \includegraphics[width=0.99\textwidth]{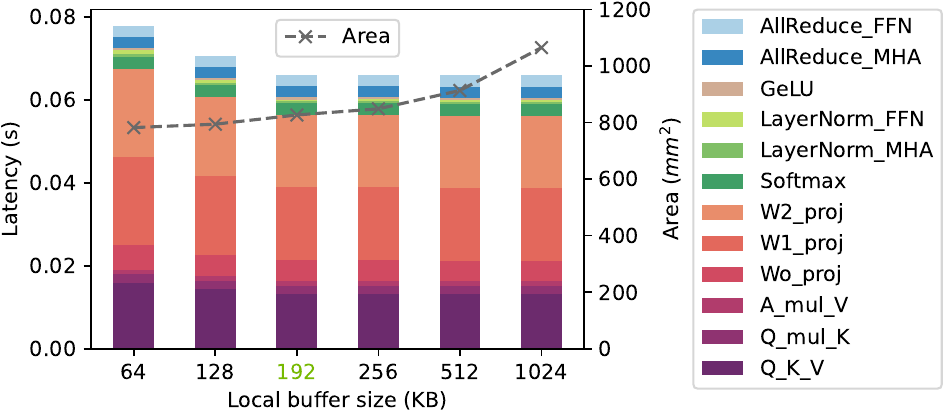}
        \caption{\textit{Prefill} Latency per GPT-3 Layer.}
        \vspace{5pt}
    \end{subfigure}
    \hfill
    \begin{subfigure}[]{0.46\textwidth} 
        \includegraphics[width=0.99\textwidth]{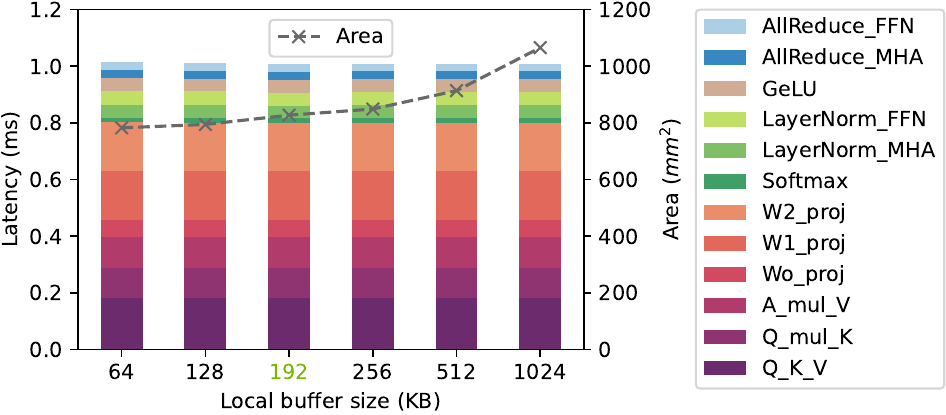}
        \caption{\textit{Decoding} Latency per GPT-3 Layer per Output Token.}
    \end{subfigure}
    
    \caption{Impact of Local Buffer Size on Performance.}
    \vspace{-5pt}
    \label{fig:local buffer}
\end{figure}

\textbf{Local Buffer}. We fix the hardware specifications to an NVIDIA \textcolor{nvidiagreen}{A100} (as in Table~\ref{tab:example}) and sweep local buffer size. The results are shown in Figure~\ref{fig:local buffer}. For \prefill{}, increasing the local buffer size from 64KB to 192KB improves the performance by 18.0\% while increasing the area by 5.8\%. Further increasing to 1024KB has a negligible performance gain of only 0.2\% at the cost of 28.8\% bigger area. For the \decoding{} stage, increasing the local buffer size from 64KB to 1024KB only increases the performance by 0.5\%.

\textbf{Analysis:} The reduced \prefill{} latency with larger local buffers is mainly because of reduced matrix multiplication latencies. A larger local buffer enables larger matrix tiles and therefore higher systolic array utilization rate. A local buffer size of 192KB is just enough for matrix multiplication of $128\times128\times128$ at FP16 with double buffering technique. It can fully utilize the $16\times 16$ systolic arrays, shedding some insight on the NVIDIA A100's design choices. 
Increasing local buffer size when the systolic array is already fully utilized leads to marginal performance gains.
For \decoding{} stage, increasing local buffer size does not help because it's IO-bound.

\textbf{Global Buffer.} The performance trends for global buffer size are similar to Figure~\ref{fig:local buffer}. Increasing the global buffer size from 10MB to 40MB speeds up \prefill{} by 11.8\% while increasing area by 9.6\%. Further increasing to 80MB only brings a performance gain of 0.01\% at the cost of 11.7\% bigger area. For \decoding{}, increasing global buffer size from 10MB to 80MB has a performance gain of only 0.7\%.

\textbf{Analysis:} Larger global buffers enable larger matrix tiles, increasing systolic array utilization and data reuse at the global buffer level. Similarly, increasing global buffer size has diminishing returns once the systolic arrays are saturated. The \decoding{} stage is not bounded by computation so it barely benefits from the larger global buffer.

\begin{tcolorbox}[rounded corners, boxrule=1pt]
\begin{itemize}
    \itemding{\ding{175}} \textit{Large buffers help \prefill{} but not \textit{decoding}.}
    \itemding{\ding{176}} \textit{Buffers should be large enough to fully utilize the systolic arrays.}
\end{itemize}
\end{tcolorbox}

%% file: text/4-proposal.tex
\section{Efficient Hardware Design with LLMCompass}\label{sec:proposal}

Ideally, efficient hardware design will optimize for both performance and cost. This section draws from the insights in Section~\ref{sec:implication} and proposes two efficient hardware designs: a latency-oriented design and a throughput-oriented design. Both of these designs aim to reduce hardware costs while maintaining or improving performance. The key specifications are shown in Table~\ref{tab:proposal}. All the other specifications (\textit{e.g.}, frequency, register file size, device-device interconnect, kernel launch overhead, and framework overhead~\textit{etc.}) are the same as an NVIDIA GA100 for fair comparison.

\begin{table}[!t]
    \centering
    \caption{Comparison with NVIDIA GA100}
    \resizebox{0.48\textwidth}{!}
    {
    \begin{tabular}{c|ccc}
    \toprule
    \multirow{2}{*}{\textbf{Specifications}} & \textbf{Latency} & {\textbf{GA100}} & \textbf{Throughput} \\
    & \textbf{Design} & \textbf{(Full)} & \textbf{Design} \\
    \midrule
    Core count&\textbf{64}&128&\textbf{64}\\
    Lane count& 4 & 4 &4\\
    Vector width & 32 & 32 & 32\\
    Systolic array & $16\times16$ & $16\times16$ & $\mathbf{32\times32}$\\
    Local buffer (KB)& 192 & 192 & \textbf{768}\\
    Global buffer (MB)& \textbf{24} & 48 & 48\\
    {Global buffer (bytes/clk)} & \textbf{2560} & {{5120}} & {{5120}}\\
    Memory bandwidth (TB/s) & 2 & 2 & \textbf{1}\\
    Memory capacity (GB) & 80 & 80 & \textbf{512}\\
    Memory protocol & HBM2E & HBM2E & \textbf{PCIE 5.0/CXL}\\
    \midrule
    Die area (TSMC 7nm, $mm^2$)& 478 & 826 & 787\\
    Normalized performance & 0.95 & 1 & 1.41\\
    Estimated die cost & \$80 & \$151 & \$142 \\
    Estimated memory cost & \$560 & \$560 & \$154 \\
    Estimated total cost & \$640 & \$711 & \$296 \\
    Normalized performance/cost & \textbf{1.06} & 1 & \textbf{3.41}\\
    \bottomrule 
    \end{tabular}
    }
    \label{tab:proposal}
    \vspace{-10pt}
\end{table}

\subsection{Latency-Oriented Design}

LLM inference latency means the total time between receiving the request and generating the last token. It is a critical metric for interactive use cases like chatbots. It is composed of \prefill{} latency, the time to process the input sequence, and \decoding{} latency, the time to generate the output sequence in an auto-regression way. Inference latency is usually dominated by \decoding{} unless the input sequence is much longer than the output sequence. \textit{Decoding} is IO-intensive and is mostly bounded by reading model parameters and KV cache.

\textbf{Observation:} As latency is mostly IO-bound, memory bandwidth is the key to reducing latency, making HBM the best choice. However, due to the capacity limit of HBM, the batch size cannot be too large: the size of the KV cache and intermediate values is proportional to batch size. Therefore, the massive compute capability is not fully utilized.

\textbf{Proposal:} We propose an efficient latency-oriented design by pruning half of the compute capability while using the same memory system as a GA100, as shown in the left of Table~\ref{tab:proposal}.

\begin{figure}[!t]
    \centering
    \includegraphics[width=0.47\textwidth]{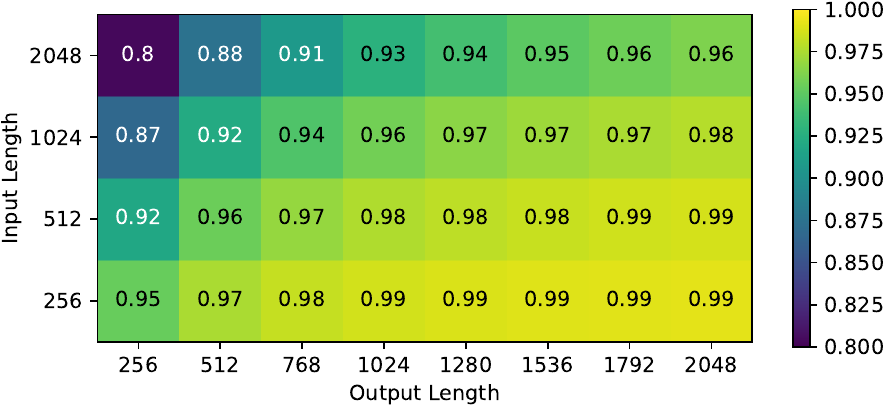}
    \caption{End-to-End Performance of Latency-Oriented Design Normalized to GA100. Performance metric: inverse of latency (higher is better). Settings: batch size\protect\footnotemark 16, 4-way tensor parallelism, running 48 GPT-3 layers (half of GPT-3).}
    \label{fig:latency design}
    \vspace{-8pt}
\end{figure}

\textbf{Results:} Compared to an NVIDIA GA100, the die area is reduced by 42.1\% while keeping 95.3\% of the performance on average. The results are shown in Figure~\ref{fig:latency design}.

\footnotetext{In reality, a batch size of 16 with input length 2048 and output length 2048 will slightly exceed the memory capacity.}

\begin{figure}[!t]
    \centering
    \includegraphics[width=0.45\textwidth]{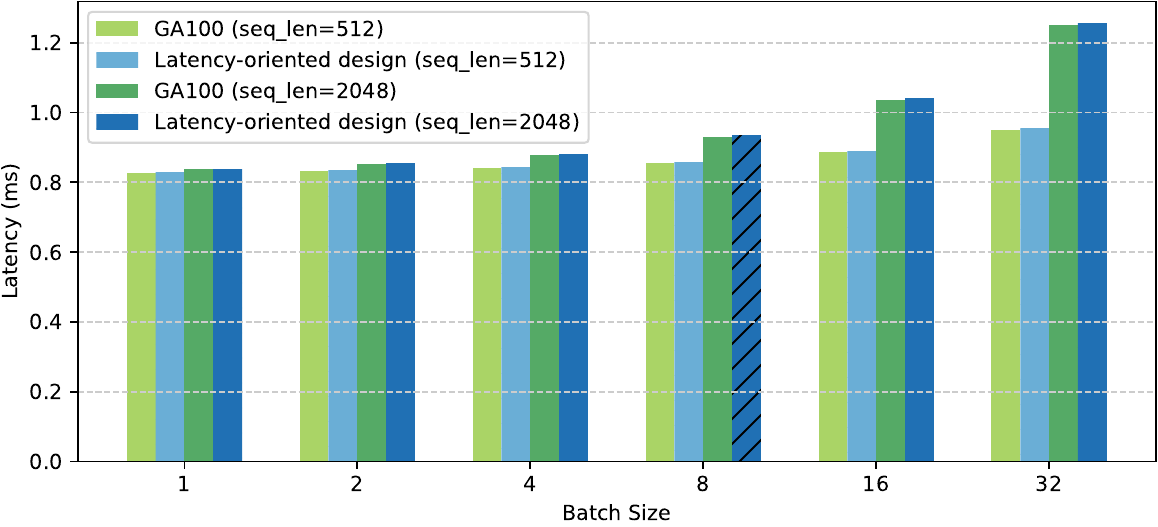}
    \caption{Comparison of \textit{Decoding} Latency per GPT-3 Layer.}
    \vspace{-8pt}
    \label{fig:latency design decoding}
\end{figure}

\textbf{Discussion:} Due to the IO-bound \decoding{} stage, the over-provisioned GA100 is not able to realize significantly improved inference performance compared to our latency-oriented design. 
As shown in Figure~\ref{fig:latency design decoding}, our pruned design achieves identical \decoding{} performance as a GA100.
The GA100 is an enormous die and is susceptible to yield issues - A100 dies are already binned to have 108 functioning SMs out of 128.
Our latency-oriented design shows that even with half the cores and SRAM disabled, the device can still achieve similar performance. This may motivate designers to salvage previously deemed faulty chips and manufacture them into separate products focused on LLM inference.

Pruning the compute capability only hurts the compute-bound \textit{prefill} performance. As \textit{prefill} is more dominant at long input sequence and short output sequence, the performance degradation will be more visible under these cases, which explains why we only achieve 80\% of the GA100 performance at input length 2048 and output length 256. With a smaller input length and larger output length, our pruned latency-aware design can achieve 99\% the performance as GA100.

\subsection{Throughput-Oriented Design}

For background use cases such as form processing or data wrangling, throughput can be more important than latency. There are generally two ways to improve throughput: 
\begin{itemize}
    \item Decrease latency - As latency is mostly IO-bound by reading parameters and KV cache, the best way to improve latency is to further improve memory bandwidth. As HBM is already expensive, this may not be easily achieved without increasing cost.
    \item Increase batch size - Generally, larger batch sizes are more efficient for throughput because the parameters are only read once for the whole batch. Larger batch sizes can also improve the hardware utilization rate. The downside is that a larger batch size consumes more compute power and increases KV cache accesses.
\end{itemize}

\textbf{Observation:} Increasing batch size is a more efficient way to improve throughput compared to decrease latency, which requires expensive high-end HBMs or even SRAMs. With a larger batch size, more memory capacity is needed to hold the larger KV cache and intermediate values.

\textbf{Proposal:} We propose a throughput-oriented design as shown in the right of Table~\ref{tab:proposal}. To hold larger batches, we use 512GB of DRAM powered by 256 PCIe 5.0 channels with an aggregated memory bandwidth of 1TB/s. (According to our area model, an 800$mm^2$ die's perimeter is able to fit around 400 PCIe 5.0 channels.) Considering the high cost and limited capacity of HBMs, this design is more cost-effective. With larger batch sizes comes a greater need for compute capability, so we quadruple the systolic arrays and the local buffer. We halve the core count and vector unit to maintain a similar die area as GA100.

\begin{figure}[!t]
    \centering
    \begin{subfigure}[]{0.45\textwidth} 
        \includegraphics[width=0.99\textwidth]{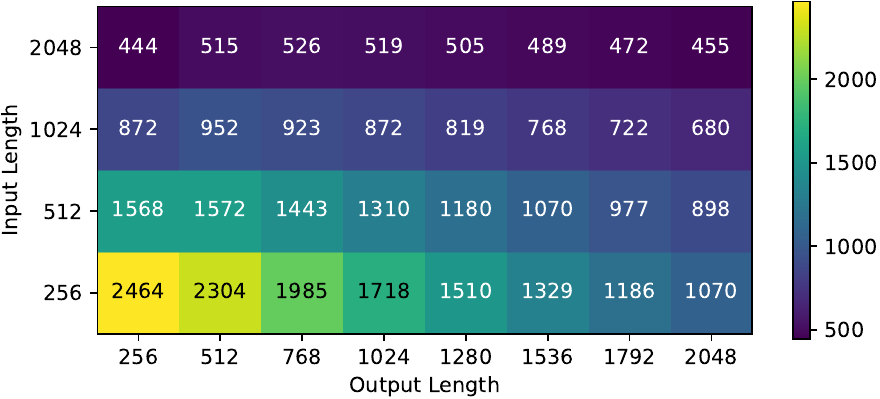}
        \caption{Throughput of Throughput-Oriented Design (Tokens/s).}
        \vspace{4pt}
        \label{fig:throughput throughput}
    \end{subfigure}
    \begin{subfigure}[]{0.45\textwidth} 
        \includegraphics[width=0.99\textwidth]{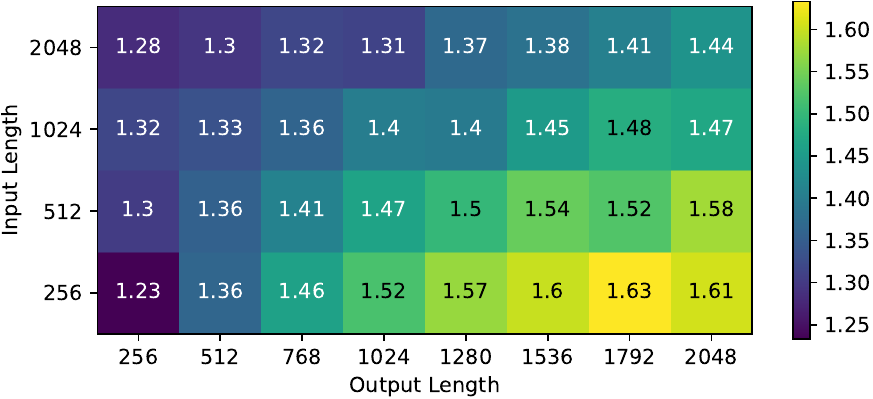}
        \vspace{-2pt}
        \caption{Normalized to a 8-GA100 GPU Node.}
    \end{subfigure}
    \caption{End-to-End Performance of Throughput-Oriented Design. Performance metric: throughput. Settings: largest batch size within memory capacity, 8-way pipeline parallelism where each device runs 12 GPT-3 layers (1/8 of GPT-3).}
    \vspace{-7pt}
    \label{fig:throughput design norm}
\end{figure}

\textbf{Results:} Compared to an NVIDIA GA100, the die area is slightly smaller and the throughput is improved by 1.42x on average. The results are shown in Figure~\ref{fig:throughput design norm}. By replacing HBMs with traditional DRAMs, the cost is reduced by 58.3\%, making a total of 3.41x gain in performance/cost.

\textbf{Discussion:} Our design has 6.4x the memory capacity of a GA100, which allows more than 12x bigger batch size after subtracting the fixed space occupied by model parameters. Ideally, with half the bandwidth of a GA100, this configuration can achieve more than 6x improvement in throughput. However, batching only reduces model parameter accesses but not KV cache reads. With a much larger batch, KV cache accesses become the new bottleneck, which diminishes the benefits of batching. As input length and output length increase, throughput decreases due to longer KV cache reads, as shown in Figure~\ref{fig:throughput throughput}.

From a latency perspective, this throughput-oriented design may not be promising: the latency is 9.21x worse than GA100 on average. While model parameters are only read once for each batch, a larger batch size means more KV cache and intermediate values to read. In LLM inference, there is no free lunch between latency and throughput.

%% file: text/7-conclusion.tex
\section{Related Work}
\subsection{Evaluating Large-scale Hardware Design}

Evaluating the various characteristics of a hardware design, including performance, area, and cost, is extremely useful for hardware designers.
To this end, the options are as follows:

\textbf{Roofline Model Analysis}~\cite{williams2009roofline}. Roofline models are analytical, fast to evaluate, and can be applied to various architectures for performance comparison. However, they can be overly optimistic relative to actual hardware capabilities.



\textbf{Cycle-level Simulation}~\cite{gpgpu-sim,gem5apu,gem5-apu2,MGPU-sim,macsim,kim2012macsim,gong2017multi2sim,ubal2012multi2sim,khairy2020accel, parashar2019timeloop,scale-sim-1,scale-sim-2}. With a typical simulation rate of less than 100K instructions per second, cycle-level simulators become infeasible for evaluating LLM scale workloads. 
As these simulators are often designed for specific architectures, it is hard to describe a hardware design very different from its design purpose (\textit{e.g.}, it's almost impossible to use GPGPU-sim~\cite{gpgpu-sim} to evaluate a TPU-like design because it relies on the GPU ISA). These simulators often require the user to provide the program for evaluation. If the software program is not optimized, it may lead to unfair comparisons.

\textbf{FPGA Emulation.} Another way is to implement the design in RTL code and emulate it on FPGAs. The RTL code can be either handwritten or generated by accelerator generators~\cite{venkatesan2019magnet,nvdla,skybox,genc_gemmini_dac2021}. Although the emulation is fast, the synthesis process may take a long time for a large design capable of running large language models, and we may need multiple FPGAs to hold the design. Additionally, users need to rewrite the RTL code and redo the synthesis to evaluate a new design. For each new design, users are responsible for mapping their workloads efficiently to fully utilize the hardware.

\textbf{Comparison.} As shown in Table~\ref{tab:evalution method comparison}, LLMCompass is more accurate than roofline model analysis, faster and more versatile than cycle-level simulators, and less engineering-intensive than FPGA emulation.
For evaluating large-scale hardware design in the era of LLMs, we believe LLMCompass is helpful for the initial design stage to determine high-level hardware characteristics (\textit{e.g.}, number of cores, memory bandwidth, \textit{etc}.).
While this work describes LLMCompass in the context of large Transformer models, it can also be applied to other dense neural network models.

\textbf{LLMCompass can complement FPGA emulation.} Designers can perform initial design space exploration before incurring the heavy costs associated with FPGA emulation and the necessary RTL implementation of the proposed design. 

\subsection{Accelerator Design Space Exploration}

Since the era of CNN, various works have focused on exploring optimal hardware designs as well as mapping~\cite{parashar2019timeloop,dave2019dmazerunner,dave2020dmazerunner,lu2017flexflow, venkatesan2019magnet, yang2020interstellar, mindmapping,li2021analytical,reagen2017case, fast}. LLMCompass is different from these works in design considerations and emphasis: \ding{172} Mainly targeting Convolutional Neural Networks (CNNs), these works focus on loop parallelization, loop order, and data flows (\textit{e.g.}, weight stationary or output stationary), which are not the primary design considerations in Transformer-based LLMs. LLMCompass is more tailored for matrix multiplication tiling and scheduling as well as other Transformer operators such as \textit{LayerNorm}. \ding{173} LLMCompass is designed for GPU-scale designs, which are much larger than CNN accelerators like Eyeriss~\cite{eyeriss}. LLM workloads are also significantly larger than CNN workloads.

\textbf{LLMCompass can also complement design space explorations.} Implemented as a Python library, LLMCompass can be seamlessly integrated into design space exploration frameworks such as FAST~\cite{fast}. FAST uses an internal TPU performance simulator, limiting its broader utility. Fast and accurate, we believe the fully open-source LLMCompass can democratize hardware design space exploration research.

\begin{table}[!t]
    \caption{Comparison of Hardware Evaluation Methods}
    \centering
    \setlength\tabcolsep{3pt}
    \resizebox{0.48\textwidth}{!}
    {
    \begin{tabular}{c|cccc@{}c}
    \toprule
    \multirow{2}{*}{\textbf{Methods}}  & \multirow{2}{*}{\textbf{Fast}} & \multirow{2}{*}{\textbf{Accurate}} & \textbf{Architecturally} & \textbf{Performance} & {\textbf{Cost}}\\
    & & & \textbf{Descriptive\ding{117}} & \textbf{Optimal\ding{118}}& \textbf{Aware}\\
    \midrule
    Roofline & \ding{51} &\ding{55} & \ding{51} & \ding{51}&\ding{55} \\
    Cycle-level & \ding{55} & \ding{51} & \ding{55}&\ding{92}&\ding{55} \\
    FPGA & \ding{92} & \ding{51} &\ding{92} &\ding{92} & \ding{51}\\
    \textbf{LLMCompass} & {\ding{51}} & \ding{51} & \ding{51} & \ding{51} & \ding{51}\\
    \bottomrule 
    \end{tabular}
    }
    \caption*{\footnotesize{ \ding{117}: The ability to describe different hardware designs.}}
    \vspace{-7pt}
    \caption*{\footnotesize{\ding{118}: Find the optimal mapping to fully demonstrate hardware capability.}}
    \vspace{-14pt}
\end{table}\label{tab:evalution method comparison}

    

\subsection{Accelerating LLM Inference}

Many Transformer accelerators have been proposed~\cite{tambe2021edgebert,transformeraccelerator,spatten,ham2021elsa}, mainly focusing on accelerating the Transformer with hardware-software co-design such as pruning or approximate-computing. Whether these techniques are effective for the largest of models remains to be seen. Additionally, the major challenge of LLMs today comes from the massive scale of the models, which is the main scope of this paper.

Many efforts have also been made to accelerate LLM inference at the software domain~\cite{deepspeed,pope2023efficiently,dao2022flashattention,dao2023flashattention,flexgen}. LLMCompass is compatible with these optimization techniques by modeling their compute and memory access patterns. We don't discuss techniques like FlashAttention~\cite{dao2022flashattention} because they are orthogonal to the focus of this paper: They focus on the software domain and are usually implemented on a specific hardware platform such as NVIDIA GPUs. 

\section{Conclusion}
\label{sec:conclusion}

This work introduces LLMCompass, a fast, accurate, and architecturally descriptive hardware evaluation framework for LLM inference workloads. 
LLMCompass' hardware description template, mapper, and architectural simulator allow hardware designers to evaluate large-scale chip designs for LLMs, which are infeasible for cycle-level simulators. The incorporated area and cost models can also help designers reason about performance-cost trade-offs.
With the aid of LLMCompass, we draw implications on how hardware designs affect LLM inference.
Based on these findings, we propose a latency-oriented design and a throughput-oriented design that achieve 1.06x and 3.41x performance per cost improvements respectively, compared to NVIDIA GA100. We plan to extend LLMCompass to support more machine learning workloads as well as LLM fine-tuning in the future.

\section*{Acknowledgements}

We would like to thank Qixuan (Maki) Yu, Zhongming Yu, Haiyue Ma, Christopher Batten, and the entire Princeton Parallel Group, for their feedback, suggestions, and encouragement. This material is based upon work supported by the National Science Foundation Graduate Research Fellowship Program under Grant No. DGE-2039656, the National Science Foundation under Grant No. CCF-1822949, Air Force Research Laboratory (AFRL) and Defense Advanced Research Projects Agency (DARPA) under agreement No. FA8650-18-2-7862. Any opinions, findings, and conclusions or recommendations expressed in this material are those of the author(s) and do not necessarily reflect the views of the National Science Foundation. The U.S. Government is authorized to reproduce and distribute reprints for Governmental purposes notwithstanding any copyright notation thereon. The views and conclusions contained herein are those of the authors and should not be interpreted as necessarily representing the official policies or endorsements, either expressed or implied, of Air Force Research Laboratory (AFRL) and Defense Advanced Research Projects Agency (DARPA) or the U.S. Government.